\documentclass{PoS}
\usepackage[T1]{fontenc}
\usepackage[utf8]{inputenc}
\usepackage{amsmath,amssymb,amsfonts}
\usepackage{multicol}
\usepackage{wrapfig}
\usepackage{defs}

\title{A status update on the determination of $\Lambda_{\MSbar}^{N_{\rm f}=3}$ \\
by the ALPHA collaboration}

\ShortTitle{ALPHA collaboration, update on $\Lambda_{\MSbar}^{N_{\rm f}=3}$}

\author{%
M.~Dalla~Brida$^{a}$, %
P.~Fritzsch$^{b}$, %
T.~Korzec$^{c,\,d}$, %
A.~Ramos$^{e}$, %
\speaker{S.~Sint}$^{\,f}$ and %
R.~Sommer$^{a}$%
\vskip0.25em\\
\llap{$^a$} NIC, DESY, Platanenallee~6, D-15738~Zeuthen, Germany\\
\llap{$^b$} Instituto de F\'{\i}sica Te{\'o}rica UAM/CSIC, Universidad Aut{\'o}noma de Madrid,\\
C/ Nicol{\'a}s Cabrera 13-15, Cantoblanco, Madrid 28049, Spain\\
\llap{$^c$}
Institut~f\"ur~Physik, Humboldt-Universit\"at~zu~Berlin, 
Newtonstr.~15, D-12489~Berlin, Germany\\
\llap{$^d$}
Department of Physics, Bergische Universit\"at Wuppertal, Gau\ss str. 20,\\
D-42119 Wuppertal, Germany\\
\llap{$^e$}
PH-TH, CERN, CH-1211 Geneva, Switzerland\\
\llap{$^f$}School of Mathematics, Trinity College Dublin, Dublin 2, Ireland
\vskip0.25em\\
E-mail:~\email{mattia.dalla.brida@desy.de},\hspace{1ex}
\email{p.fritzsch@csic.es},\hspace{1ex}
\email{korzec@physik.hu-berlin.de},\hspace{1ex} 
\email{alberto.ramos@cern.ch},\hspace{1ex}
\email{sint@maths.tcd.ie},\hspace{1ex}
\email{rainer.sommer@desy.de} %
}

\abstract{The ALPHA collaboration aims to determine $\alpha_s(m_Z)$ with a total error
below the percent level. A further step towards this goal can be taken by combining
results from the recent simulations of $2+1$-flavour QCD by the CLS initiative with a number of tools
developed over the years: renormalized couplings in finite volume schemes, recursive finite size techniques,
two-loop renormalized perturbation theory and the (improved) gradient flow on the lattice.
We sketch the strategy, which involves both the standard SF coupling in the high energy regime
and a gradient flow coupling at low energies. This implies the need for matching both schemes at
an intermediate switching scale, $\Lswi$, which we choose roughly in the range 2-4~$\GeV$.
In this contribution we present a preliminary result for this matching procedure, and  
we then focus on our almost final results for the scale evolution of the SF coupling 
from $\Lswi$ towards the perturbative regime, where we extract the $\Nf=3$ $\Lambda$-parameter,
$\Lambda_{\MSbar}^{(3)}$, in units of $\Lswi$.
Connecting $\Lswi$ and thus the $\Lambda$-parameter to a hadronic scale such as $F_K$ requires 2 further ingredients:
first, the connection of $\Lswi$ to $\Lmax$ using a few steps with the step-scaling function
of the gradient flow coupling, and, second, the continuum extrapolation of $\Lmax F_K$.

\begin{flushleft}
{\hfill\parbox{3.8cm}{\small%
TCDMATH 15--11\\
DESY 15-218\\
CERN-PH-TH-2015-269\\
WUB/15-08\\
HU-EP-15/54\\
IFT-UAM/CSIC-15-122}}
\end{flushleft}

}

\FullConference{The 33rd International Symposium on Lattice Field Theory\\
		14 -18 July 2015\\
		Kobe International Conference Center, Kobe, Japan}

\begin{document}

\section{Introduction}
One of the long-term goals of the ALPHA-collaboration consists in a controlled and precise
determination of the strong coupling constant, $\alpha_s$,
from experimentally well-measured hadronic observables. In 2+1-flavour QCD
the hadronic input can be obtained using gauge field ensembles produced by
Coordinated Lattice Simulations (CLS) effort~\cite{CLS, StefanSchaeferlat15}.
The 3 bare parameters, $m_u=m_d$, $m_s$ and the coupling $g_0$ can be fixed
taking e.g.~$m_\pi$, $m_K$ and $F_K$ from experiment, after
correcting for isospin breaking and electromagnetic effects.
Once this is achieved, any other observable of the theory becomes a prediction,
in particular, the renormalized strong coupling at a large scale,
\begin{equation}
  \alpha_s^{(\Nf=3)}(1000\times F_K),\qquad \alpha_s = \dfrac{\bar{g}^2}{4\pi}, 
\end{equation}
in any renormalization scheme. Obviously the final goal is to determine $\alpha_s^{(\Nf=5)}(m_Z)$ 
in the $\MSbar$-scheme of dimensional regularization. This requires a matching procedure to
$4$- and $5$-flavour QCD across the charm and bottom thresholds, respectively.
There is some hope that perturbation theory may be adequate to
include not only the effects of the heavier bottom quark on the coupling,
but also those of the charm quark~\cite{Bruno:2014ufa,Grozin:2011nk}. This will not be discussed here.
We just mention that current phenomenological estimates of $\alpha_s$ yield
\begin{equation}
  \alpha_s(m_Z) = 0.1183(12)\,,
\end{equation}
with a 1\% error~\cite{PDG}. This is compatible with the currently available lattice results
which have total errors of similar size~\cite{FLAG:2013}. Using the strategy described
in the following we expect a significant reduction of this error.

\section{The $\Lambda$-parameter and $\alpha_s$}

Given the strong coupling $\alpha_s = \bar{g}^2/(4\pi)$ in a quark mass independent renormalization scheme
the corresponding  $\Lambda$-parameter is defined by
\begin{equation}
    \Lambda^{(\Nf)}= \mu \big[b_0\bar{g}^2(\mu)\big]^{-b_1/(2b_0^2)}\,\rme^{-1/(2b_0\bar{g}^2(\mu))}
                       \exp\bigg\{\!\!-\!\!\int_{0}^{\bar{g}^{\vphantom{A}}(\mu)}\! dg
                       \bigg[\dfrac{1}{\beta(g)}+\dfrac{1}{b_0g^3} -\dfrac{b_1}{b_0^2g}\bigg]\bigg\},
\label{eq:Lambda}
\end{equation}
where $\Nf$ denotes the fixed number of quark flavours.
If the coupling constant is non-perturbatively defined,
so is the $\beta$-function and thus the $\Lambda$-parameter. Nevertheless,
at high scales the perturbative form $\beta(g) = -b_0 g^3-b_1 g^5 +\ldots$ can
be used, with the universal coefficients,
\begin{equation}
   b_0= \dfrac{11-\frac23 \Nf}{(4\pi)^2},\qquad
   b_1= \dfrac{102-\frac{38}{3} \Nf}{(4\pi)^4}\,.
\end{equation}
The relation between the $\Lambda$-parameters in any two schemes is {\em exactly} 
calculable by one-loop perturbation theory: for
schemes ${\rm X}$ and ${\rm Y}$ the one-loop relation between the respective couplings
entails the relation,
\begin{equation}
    g_{\rm X}^2 = g_{\rm Y}^2 + c_{\rm XY}^{} g_{\rm Y}^4 + \rmO(g_{\rm Y}^6)
    \quad \Rightarrow \quad
    \dfrac{\Lambda_{\rm X}}{\Lambda_{\rm Y}} = {\rm e}^{c_{\rm XY}^{}/2b_0}\,.
\end{equation}
Hence, even though the $\overline{\rm MS}$-scheme is only defined perturbatively,
its $\Lambda$-parameter can be defined non-perturatively and is conventionally used
as a reference. Note that a 1\% error on $\alpha_s(m_Z)$ translates to a 6-7\% error on
$\Lambda^{(5)}_{\MSbar}$~\cite{PDG}, which sets the reference in terms of the precision.

\section{An approach involving 2 couplings}

\begin{figure}[h]
     \hspace{2cm}
	\setlength\fboxsep{0pt}
        \setlength\fboxrule{0.0pt}
        \fbox{\includegraphics[trim = 0mm 0mm 50mm 150mm, clip,width=0.7\textwidth]{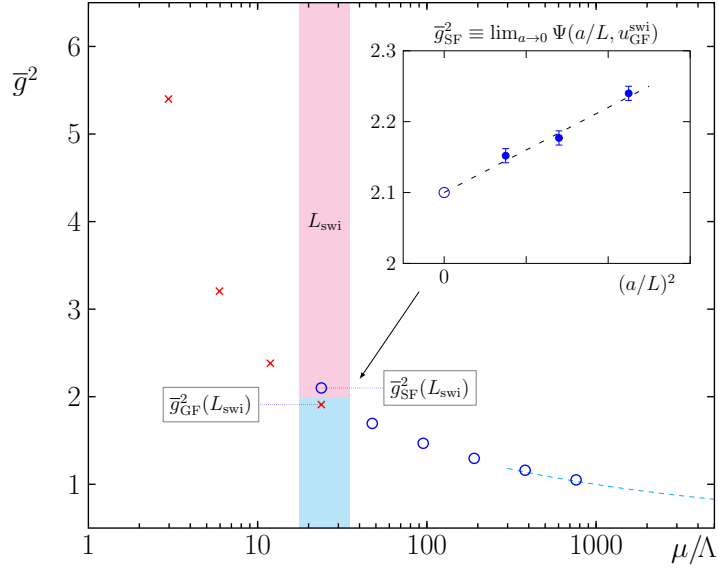}}
  \caption{A sketch of the strategy (cp.~text), reproduced from ref.~\cite{Brida:2014joa}.
          }
\label{fig:strategy}
\end{figure}

The approach  to compute $\Lambda^{(3)}_{\MSbar}$ followed by our collaboration
has previously been described in ref.~\cite{Brida:2014joa} and is sketched in fig.~\ref{fig:strategy}.
Moving up the energy scale from the hadronic to the perturbative regimes it consists of the four separate steps:
\begin{enumerate}
\item Matching of a hadronic scale, e.g.~$F_K$, to the box size $\Lmax$ defined
implicitly through the gradient flow coupling of ref.~\cite{Fritzsch:2013je},
$\bar{g}^2_{\rm GF}(\Lmax) = u_{\rm max}$. Here the value $u_{\rm max}$ must be
chosen such that there is a range of common values of $\beta=6/g_0^2$, so that
the limit
\begin{equation}   
   \lim_{a \rightarrow 0} \left(a F_K\right)(\beta) \times \left(L_{\rm max}/a\right)(\beta)\,,
\end{equation}
can be taken. As the matching proceeds via the bare parameters
the same lattice action must be used as in the CLS simulations.
\item
As a next step we calculate the evolution of the GF coupling from $\Lmax$ to
an intermeditate scale, $\Lswi=2^{-3}\Lmax$, by 3 steps up the energy scale by factors
of 2. One may do a step more or less, and the scale factor need not be 2 in all the steps.
In fact, $\Lswi$  can be defined in a variety of ways. The only thing that matters is that the relation
between $\Lmax$ (and thus $F_K$) and a given choice of $\Lswi$ is known precisely.

\item At the intermediate scale $L_{\rm swi}$ we switch scheme to the
traditional SF coupling~\cite{Luscher:1992an,Sint:1995ch}, $\bar{g}^2_{\rm SF}(\Lswi)$. At this point we
also change the gauge action to the Wilson plaquette action.

\item The evolution of the SF coupling is traced by doing 3-5 steps up the energy scale
with scale factor 2. At small couplings the $\Lambda$-parameter can then
be extracted in units of $\Lswi$ 
using a perturbative evaluation of the exponent in eq.~(\ref{eq:Lambda}).
\end{enumerate}
We  use 2 different couplings and gauge actions for technical reasons: first, the
target precision for the $\Lambda$-parameter from eq.~(\ref{eq:Lambda}) requires
the knowledge of the 3-loop $\beta$-function, which is only available for the SF coupling~\cite{Bode:1999sm}.
Furthermore, cutoff effects in the SF coupling have been calculated to 2-loop order
with the Wilson gauge action and their subtraction from the data, 
helps to stabilise the continuum limit. Finally the scaling properties of the statistical errors
at fixed $L/a$ with the scale $L$ favour the SF coupling towards the perturbative high
energy regime~\cite{Brida:2014joa}. Combining the advantages of both couplings
seems the best strategy, the price to pay being the additional scheme switching step.

\section{Computation of $\Lswi\Lambda$ }

\begin{figure}[t]
        \vspace{-1.5em}
        \setlength\fboxsep{0pt}
        \setlength\fboxrule{0.0pt}
        \fbox{\includegraphics[trim = 0mm 0mm  0mm 0mm, clip,width=0.49\textwidth]{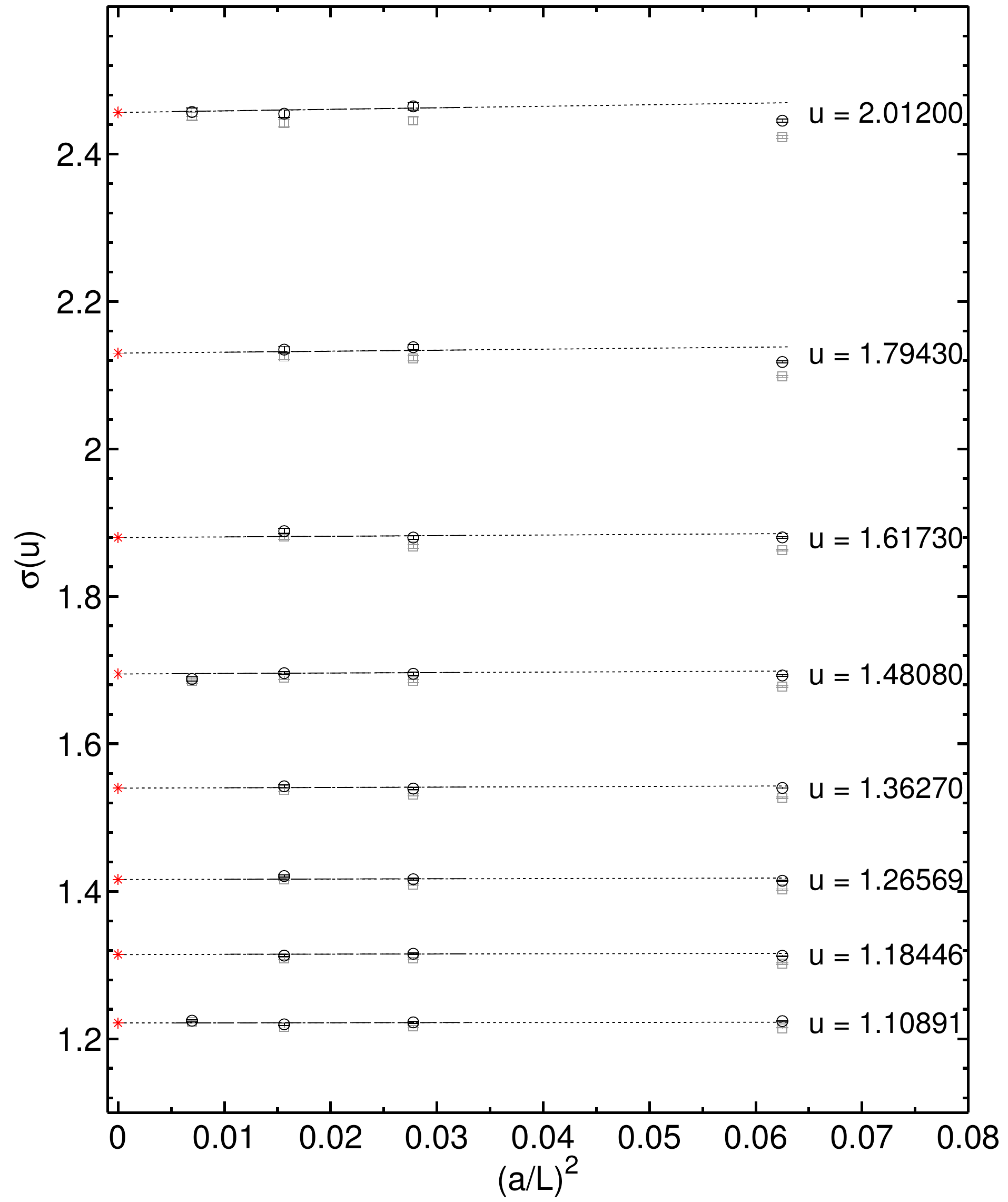}}
        \fbox{\includegraphics[trim = 0mm 0mm  0mm 0mm, clip,width=0.49\textwidth]{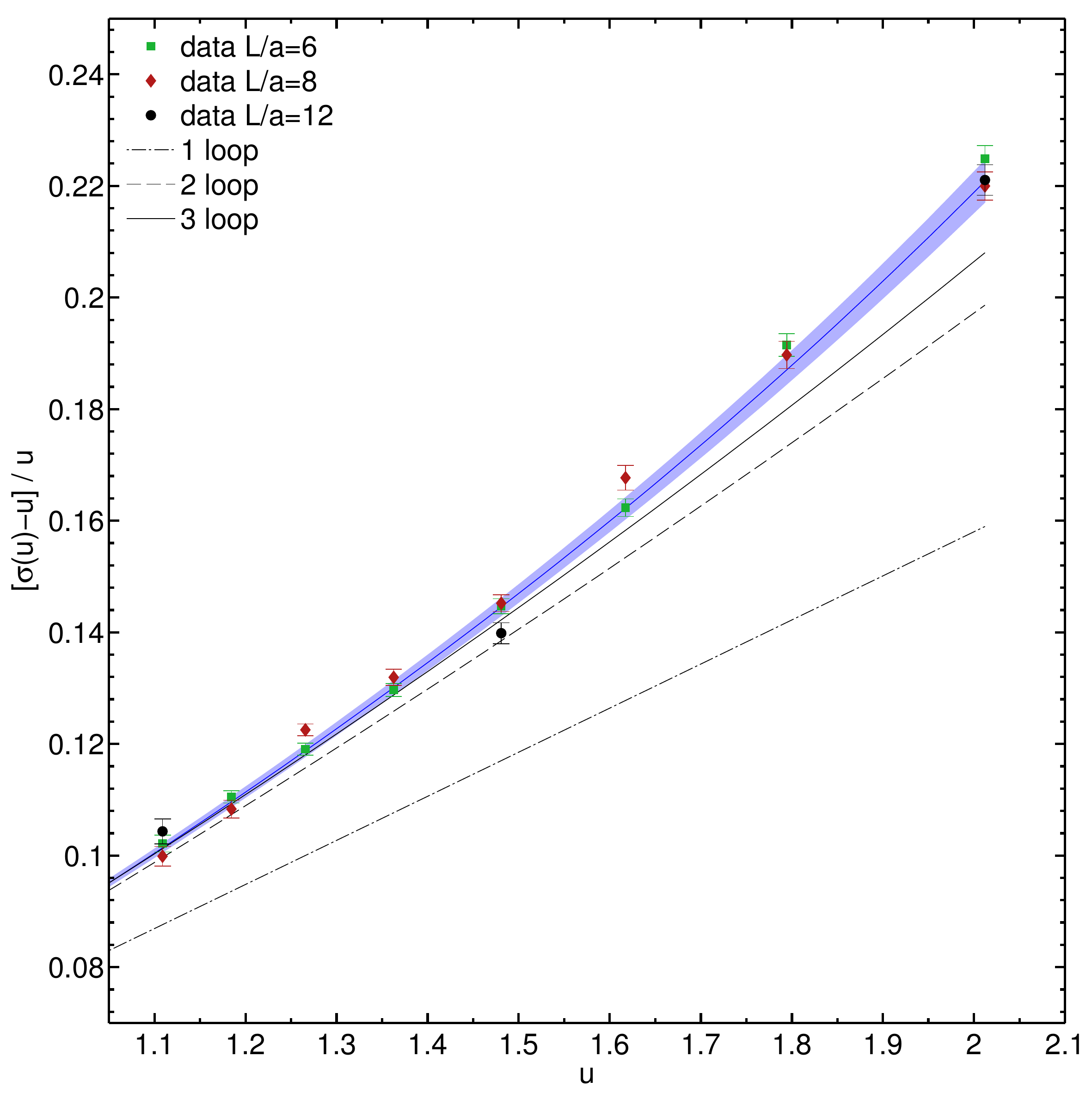}}
        \vspace{-0.0cm}
\caption{
{\em Left panel}: The data points interpolated to common $u$-values, together with the global fit function,
The rightmost points were not included in the fit.
 {\em Right panel}: The continuum extrapolated SSF with its error band (shaded area) together with the
data points (cp.~text).}
\label{fig:SSF}
\end{figure}
\begin{wrapfigure}{r}{0.5\textwidth}
  \vspace{-1em}
  \begin{center}
        \setlength\fboxsep{0pt}
        \setlength\fboxrule{0.0pt}
        \fbox{\includegraphics[trim = 0mm 0mm 0mm 0mm, clip,width=0.5\textwidth]{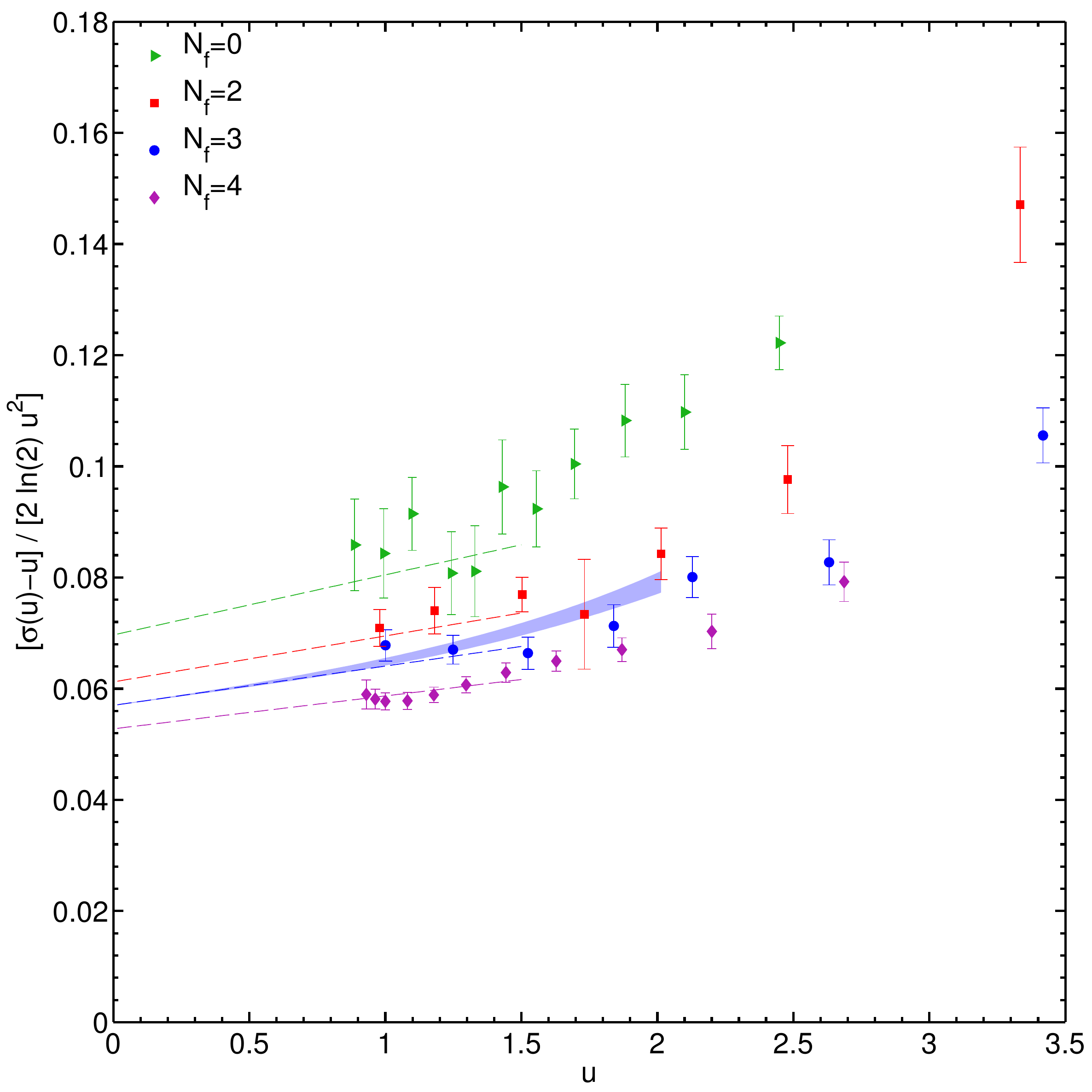}}
  \end{center}
  \vspace{-1em}
  \caption{Comparison with results from the literature for $\Nf=0,2,3,4$
  \cite{Luscher:1993gh,
  DellaMorte:2004bc,Fritzsch:2012wq,Aoki:2009tf,
  Tekin:2010mm}.
  The shaded area corresponds to the global fit. The dashed
  lines show the perturbative prediction $b_0+\rmO(u)$.}
  \vspace{1ex}
\label{fig:Nfdep}
\end{wrapfigure}

The scale evolution of the SF coupling $\bar{g}^2(L)$ can be traced by constructing
the step-scaling function (SSF)
\begin{equation}
  \sigma(u) = \left.\bar{g}^2(2L)\right\vert_{\bar{g}^2(L)=u},
\end{equation} 
as the continuum limit of lattice approximants, $\Sigma(u,a/L)$.
This requires to compute the SF coupling on pairs of lattices
$L/a$ and $2L/a$ at the same bare parameters.
We have measured $\Sigma(u,L/a)$ for lattice sizes $L/a=4,6,8$ and in 3 cases for $L/a=12$,
and for 6 approximately tuned $u$-values in the interval $[1.08,2.012]$.
The critical mass has been tuned to high precision so that the
corresponding systematic effects are completely negligible.
Furthermore, the knowledge of perturbation theory to 2-loop order allows
to subtract the cutoff effects completely at a given perturbative order:
with
\begin{equation}
  \delta(u,a/L) = \dfrac{\Sigma(u,a/L)}{\sigma(u)}-1  = \delta_1(L/a)\times u + \delta_2(L/a)\times u^2 + \rmO(u^3),
\end{equation}
we may define the 2-loop perturbatively improved data,
\begin{equation}
  \widetilde{\Sigma}(u,a/L) = \dfrac{\Sigma(u,a/L)}{1+\delta_1(L/a)u + \delta_2(L/a)u^2}\,.
\end{equation}
Given the data points we perform global fits to both the perturbatively improved
and unimproved data. An example for such a fit ansatz is
\begin{equation}
  \widetilde{\Sigma}(u,a/L) = u + s_0 u^2 + s_1 u^3 + c_1 u^4 + c_2 u^5 + \rho_1 \dfrac{a^2}{L^2}  u^4,
 \label{eq:fit}
\end{equation}
which we apply to all the data with $L/a\ge 6$. Here,
\begin{equation}
  s_0= 2b_0\ln 2,\qquad s_1 = s_0^2 + 2b_1\ln 2,
\end{equation}
are fixed to their perturbative values, the fit parameters $c_1$ and $c_2$ 
describe the continuum SSF and $\rho_1$ models the cutoff effects. The main assumption
of such a global fit ansatz is the smoothness of both the continuum SSF and the cutoff effects as 
a function of $u$.  With 19 data points and 3 parameters the fit in eq.~(\ref{eq:fit}) 
has a reasonable $\chi^2/{\rm d.o.f.}=1.0$. The data points, slightly interpolated to
common values of $u$, are displayed together with the resulting  
continuum SSF in figure~\ref{fig:SSF}. A comparison with data from the literature
for $\Nf=0$~\cite{Luscher:1993gh}, $\Nf=2$~\cite{DellaMorte:2004bc,Fritzsch:2012wq}, $\Nf=3$~\cite{Aoki:2009tf}
and $\Nf=4$~\cite{Tekin:2010mm} is made in figure~\ref{fig:Nfdep}.\\
We now define the switching scale, $\Lswi$, through the SF coupling by setting,
\begin{equation}
   \bar{g}^2_{\rm SF}(\Lswi)=2.012,
\end{equation}
which is the largest $u$-value for which we have computed the SSF.
Taking this value as our starting point, we may now recursively step up the 
energy scale by factors of 2, using the fitted continnum SSF, viz.
\begin{equation}
  u_0 =\bar{g}^2_{\rm SF}(\Lswi),\quad u_i= \sigma(u_{i+1}),
\end{equation}
where $i=0,1,2,\ldots$.
Then, using the formula for the $\Lambda$-parameter, eq.~(\ref{eq:Lambda}) and its relation
to $\Lambda_{\MSbar}$~\cite{Sint:1995ch},
\begin{equation}
 \Lambda_{\rm SF}^{(3)} = 0.38286(2)\times\Lambda_{\MSbar}^{(3)}\,,
\end{equation}
we obtain stable results already after a couple of steps. As our preliminary result
we quote
\begin{equation}
 \Lswi  \Lambda_{\MSbar}^{(3)} = 0.0802(24)\,,
\end{equation}
with a total error of  3 percent.

\section{Matching to the GF coupling}

Having finished step 4 of our strategy, we now discuss step 3, the 
matching to the gradient flow coupling. We define the GF coupling in 
a finite volume with SF boundary conditions following \cite{Fritzsch:2013je}.
However, in order to reduce boundary effects, 
we restrict the sum to the chromo-magnetic components by summing
over spatial Lorentz indices only,
\begin{equation}
  -\frac12 \sum_{k,l=1}^3 t^2 \left.\langle G_{kl}(t,x)G_{kl}(t,x) \rangle
    \right\vert_{x_0=T/2,T=L,t=c^2 L^2/8,m_{\rm q}=0} = {\cal N}(c,a/L) \times \bar{g}_{\rm GF}^2(L)\,.
\end{equation}
We use both the Wilson flow~\cite{Narayanan:2006rf,Luscher:2010iy} and
the O($a^2$) improved Zeuthen flow~\cite{Ramos:2015baa,Ramos:2014kka} and an O($a^2$) improved definition
of the observable~\cite{Ramos:2015baa,Fodor:2014cpa}. For the matching we employ the Wilson plaquette action,
whereas the scale evolution (step 2 of the strategy)
will use the L\"uscher-Weisz gauge action to match the set-up of CLS.
\begin{figure}[t]
        \vspace{-1cm}
        \setlength\fboxsep{0pt}
        \setlength\fboxrule{0.0pt}
        \fbox{\includegraphics[trim = 10mm 10mm 60mm 160mm, clip,width=0.9\textwidth]{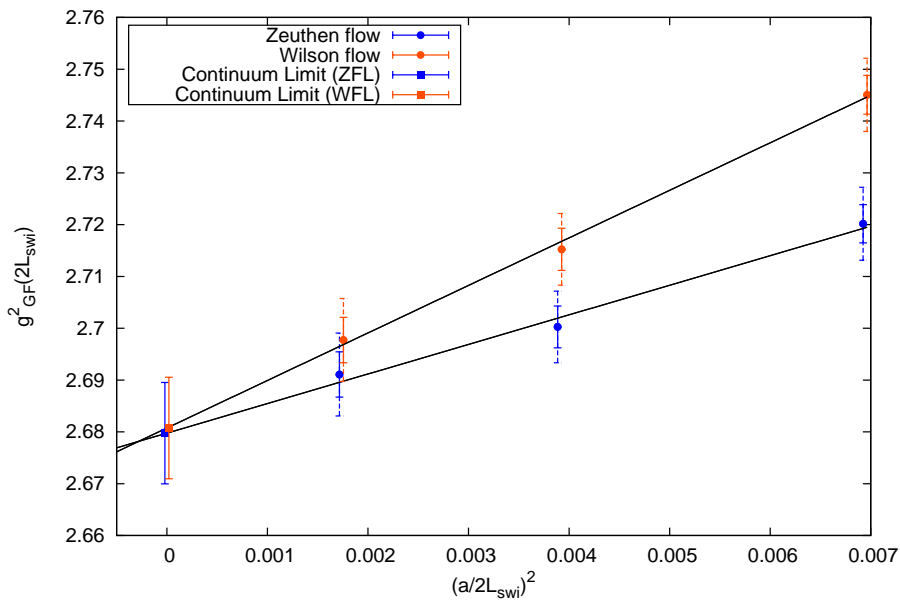}}
        \vspace{-0.5cm}
\caption{Matching the GF and SF couplings using the line of constant physics $\bar{g}_{\rm SF}^2(\Lswi)=2.012$
(cp. text).}
\label{fig:match}
\end{figure}
We note that $c$ defines the smoothing range in units of $T=L$. Staying in the middle
of the volume, $x_0=T/2$ and setting $c=0.3$ we expect that the influence from
the time boundaries is negligible. This is further corroborated by a perturbative
analysis. Then, using the $\beta$- and $\kappa$-values that
correspond to the line of constant physics (LCP) condition $\bar{g}^2(\Lswi)=2.012$,
for $L/a=6,12,16$ we double the lattice size, switch off the background field and
thus obtain the (preliminary) results in figure~\ref{fig:match}.
As our preliminary continuum extrapolation we quote,
\begin{equation}
 \bar{g}_{\rm GF}^2(2\Lswi) = 2.680(10)\,.
\end{equation}
This illustrates the typical precision that can be obtained: note
that this error includes an estimate of the uncertainty in the definition
of the LCP (hence the 2 kinds of error bars in fig.~\ref{fig:match}).
We also mention that the cutoff effects are naturally expressed in units
of the smoothing radius, $a/\sqrt{8t}=a/(cL)$, which is the reason for doubling
the lattice size as compared to the SF coupling.

\section{Summary}

We have presented our preliminary result $\Lambda_{\MSbar}^{(3)} =0.0802(24)/\Lswi$,
where $\Lswi$ is an intermediate scale, defined implicitly by $\bar{g}_{\rm SF}^2(\Lswi)=2.012$.
Furthermore, we have matched the SF and GF couplings with the (preliminary)
result $\bar{g}_{\rm GF}^2(2\Lswi) = 2.680(10)$.
To finish the project, it remains to relate $\Lswi$ to a hadronic scale by computing
the SSF for the GF coupling towards lower enery scales,
and by matching the maximal box size, $\Lmax$, reached in this way to a hadronic scale such as $F_K$, 
as measured on the CLS gauge configurations.

\par
\begin{center}
{\bf Acknowledgments}
\end{center}
This project is part of the ALPHA collaboration research programme. We would like to thank
the computer centres at HLRN and NIC at DESY-Zeuthen for providing computing resources.
S. Sint acknowledges support by SFI under grant 11/RFP/PHY3218.


\end{document}